\documentclass{ws-jai}
\usepackage[flushleft]{threeparttable}
\usepackage{subfigure} 

\newcommand{\farcs}{\mbox{$.\!\!^{\prime\prime}$}}
\newcommand{\farcm}{\mbox{$.\!\!^{\prime}$}}

\begin{document}

\catchline{}{}{}{}{} 

\markboth{Baug et al.}{TIRCAM2 on the DOT}

\title{TIFR Near Infrared Imaging Camera-II on the 3.6-m Devasthal Optical Telescope}

\author{T. Baug$^{1,2}$, D.K. Ojha$^{2}$, S.K. Ghosh$^{2,3}$, S. Sharma$^{1}$, A.K. Pandey$^{1}$, Brijesh Kumar$^{1}$,
Arpan Ghosh$^{1}$, J.P. Ninan$^{2,4}$,\\ M.B. Naik$^{2}$, S.L.A. D'Costa$^{2}$, S.S. Poojary$^{2}$, 
P.R. Sandimani$^{2}$, H. Shah$^{2}$, B. Krishna Reddy$^{1}$, S.B. Pandey$^{1}$, and H. Chand$^{1}$}

\address{
$^{1}$Aryabhatta Research Institute of Observational Sciences, Manora Peak, Nainital 263 001, India; tapas.polo@gmail.com\\
$^{2}$Tata Institute of Fundamental Research, Homi Bhabha Road, Colaba, Mumbai 400 005, India \\
$^{3}$National Centre for Radio Astrophysics, Tata Institute of Fundamental Research, Pune 411 007, India \\
$^{4}$Department of Astronomy and Astrophysics, The Pennsylvania State University, University Park, USA
}

\maketitle

\corres{$^{1}$Corresponding author.}

\begin{history}
\received{(to be inserted by publisher)};
\revised{(to be inserted by publisher)};
\accepted{(to be inserted by publisher)};
\end{history}

\begin{abstract}
TIFR Near Infrared Imaging Camera-II is a closed-cycle Helium cryo-cooled imaging camera equipped with a Raytheon
 512\,$\times$\,512 pixels InSb Aladdin III Quadrant focal plane array having sensitivity to photons in the 1--5
 $\mu m$ wavelength band. In this paper, we present the performance of the camera on the newly installed 3.6-m
 Devasthal Optical Telescope (DOT) based on the calibration observations carried out during 2017 May 11--14 and 2017
 October 7--31. After the preliminary characterization, the camera has been released to the Indian and Belgian
 astronomical community for science observations since 2017 May. The camera offers a field-of-view of
 $\sim$86\farcs5\,$\times$\,86\farcs5 on the DOT with a pixel scale of 0\farcs169. The {\it seeing} at the telescope
 site in the near-infrared bands is typically sub-arcsecond with the best {\it seeing} of $\sim$0\farcs45 realized in
 the near-infrared $K$-band on 2017 October 16. The camera is found to be capable of deep observations in the $J$, $H$
 and $K$ bands comparable to other 4-m class telescopes available world-wide. Another highlight of this camera is the
 observational capability for sources up to Wide-field Infrared Survey Explorer (WISE) W1-band (3.4 $\mu m$) magnitudes
 of 9.2 in the narrow $L-$band ($nbL$; $\lambda_{cen}\sim$~3.59~$\mu m$). Hence, the camera could be a good complementary
 instrument to observe the bright $nbL$-band sources that are saturated in the {\it Spitzer}-Infrared Array Camera
 ([3.6] $\lesssim$ 7.92 mag) and the WISE W1-band ([3.4] $\lesssim$ 8.1 mag).
 Sources with strong polycyclic aromatic hydrocarbon (PAH) emission at 3.3 $\mu m$ are also detected. Details of the
 observations and estimated parameters are presented in this paper.
\end{abstract}

\keywords{instrumentation: detectors, instrumentation: photometers, methods: observational}

\section{Introduction} \label{sec:intro}
The Infrared Astronomy group of Tata Institute of Fundamental Research (TIFR), Mumbai, India is continuously
 involved in developing ground-based and space-based near- and far-infrared photometers/imagers and spectrographs.
 The TIFR near infrared imaging camera-I (TIRCAM1) having a 58\,$\times$\,62 pixels InSb focal plane array (FPA),
 was the first array camera developed by the group. Using the TIRCAM1, observations were mainly carried out
 during the period 2001--2006 at the focal plane of the 1.2-m Mount Abu infrared telescope (f/13 Cassegrain
 focus) of Physical Research Laboratory, India. More details about this camera and the related works can be
 found elsewhere in \citet{ghosh93}, \citet{ghosh05} and \citet{ojha02, ojha03, ojha06}.
 
Later in 2012, the camera was upgraded with a larger Raytheon 512\,$\times$\,512 pixels InSb Aladdin III Quadrant FPA.
 The optics was re-designed for use of the camera at the focal plane of the 2-m Himalayan {\it Chandra}
 Telescope (HCT; f/9) operated by Indian Institute of Astrophysics, India\footnote{https://www.iiap.res.in/iao\_telescope},
 and the 2-m Girawali Telescope (f/10) operated by Inter-University Centre for Astronomy and Astrophysics,
 India\footnote{http://igo.iucaa.in/}. The camera was therefore renamed as TIFR Near Infrared Imaging Camera-II (TIRCAM2).
 The detector is cooled to an operating temperature of 35 K by a closed-cycle Helium cryo-cooler. The InSb array in
 the TIRCAM2 is sensitive to photons in the 1--5 $\mu m$ wavelength band. However, the optics in the TIRCAM2
 restricts the operating wavelength range below $\sim$3.8 $\mu m$. The camera accommodates seven selectable
 standard near-infrared (NIR) filters (see Table~\ref{table1}) for imaging observations.

\begin{figure*}
\centering
\includegraphics[width=0.9\textwidth]{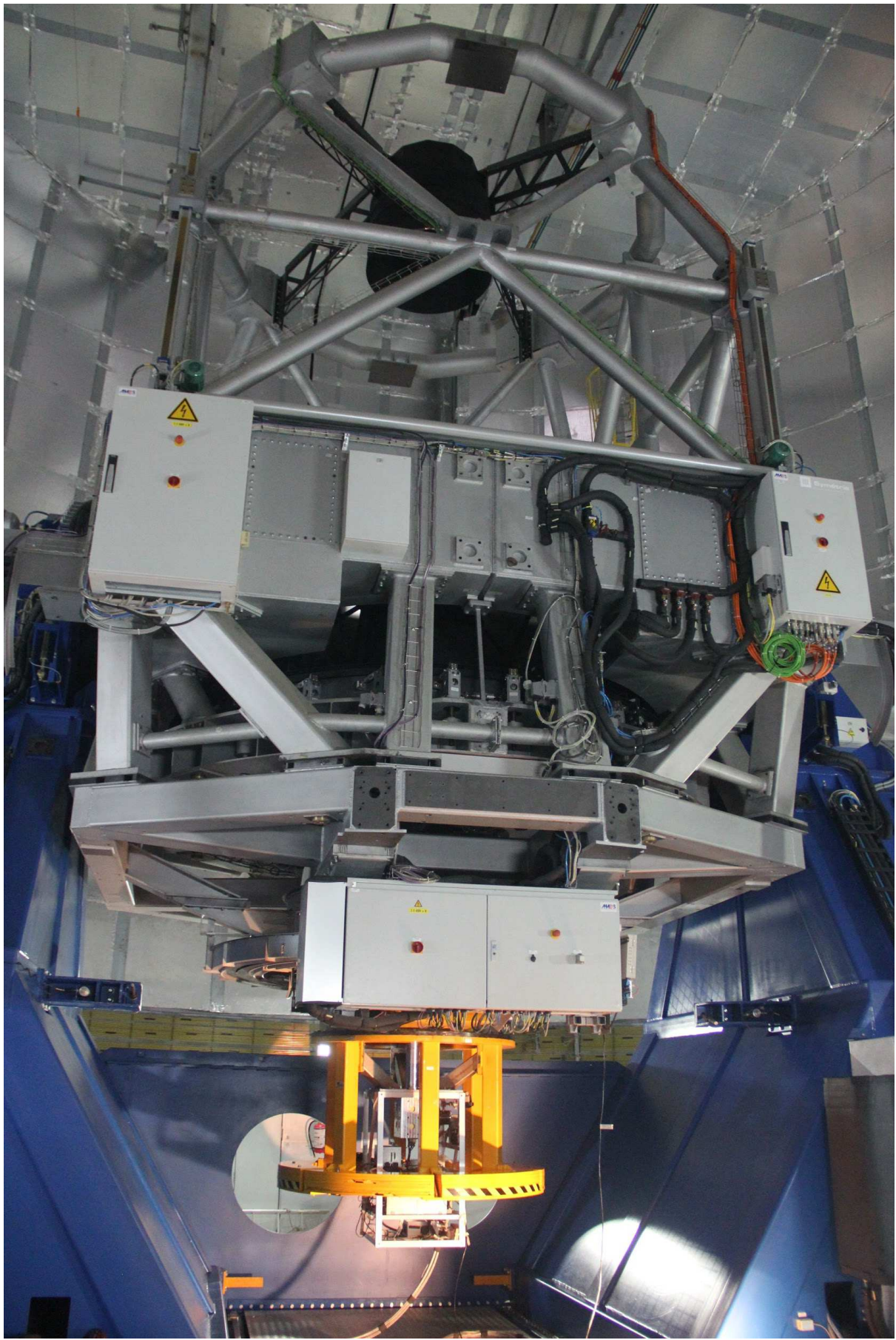}
\caption{The TIRCAM2 mounted at the axial port of the 3.6-m DOT.}
\label{fig1}
\end{figure*}

The TIRCAM2 has been recently mounted at the axial port (shown in Figure~\ref{fig1}) of the 3.6-m Devasthal Optical
 Telescope (DOT; latitude: 29$^\circ$.1971 N, longitude: 79$^\circ$.6841 E, altitude: 2450 m). The field-of-view (FoV)
 of the camera on the DOT is 86\farcs5\,$\times$\,86\farcs5, with a pixel scale of 0\farcs169$\pm$0\farcs002.
 In typical 1$''$ {\it seeing} conditions, the TIRCAM2 therefore heavily samples the stellar profile. Such a pixel
 sampling is ideal for high accuracy photometry, particularly of bright NIR sources. The dark current measured was
 $\sim$12 e$^{-}$ sec$^{-1}$, and the readout noise was $\sim$30 e$^{-}$ (single readout) for the FPA.
 This value of dark current is comparatively higher, possibly because of a thermal leak.
 The median gain of the detector was found to be $\sim$10 e$^{-}$ ADU$^{-1}$. Elaborative technical details of the
 TIRCAM2 can be found in \citet{naik12}. One of the filters (narrow $H_2$-band filter; $\lambda_{cen}\sim$ 2.12
 $\mu m$, $\Delta\lambda\sim$ 0.03 $\mu m$) in the camera, has been replaced by the standard broad $H$-band
 ($\lambda_{cen}\sim$ 1.60 $\mu m$, $\Delta\lambda\sim$ 0.30 $\mu m$) filter (see Table~\ref{table1}) before
 commissioning it on the DOT. The TIRCAM2 also accommodates the Bracket-gamma filter ($\lambda_{cen}\sim$
 2.16 $\mu m$, $\Delta\lambda\sim$ 0.03 $\mu m$). This particular filter is extremely useful in the study of
 emission line sources, and star-forming regions.
 
In this article, we report on the calibration and performance of the TIRCAM2 at the focal plane of the 3.6-m DOT, the
 largest optical aperture in the country till date. The calibration observations presented in this paper were
 mainly carried out in two successive cycles, before and after the monsoon in the year 2017 (i.e., Cycles 2017A and 2017B).
 The paper is arranged in the following manner. The details of the observations and the data reduction are presented
 in Section~\ref{obs}. The performance of the TIRCAM2, estimation of several observational parameters, and
 characterization of the camera on the DOT are presented in Section~\ref{char}. Finally, a summary of the performance
 of the TIRCAM2 is presented in Section~\ref{summary}.

\section{Observations and Data reduction}\label{obs}
The TIRCAM2 was shipped to Devasthal for installation and commissioning with the 3.6-m DOT in 2016 May. The camera
 was installed on the telescope on 2016 June 1. On the following night, 2016 June 2, first light was obtained with
 the TIRCAM2 instrument on the DOT. The observations during 2016 June could not be carried out due to the monsoons and
 cloudy weather. Preliminary observations using the TIRCAM2 with the 3.6-m DOT were performed only during 2017 January
 9--16. However, we were unable to obtain systematic data required for calibration, during that run again due to
 interruption by the clouds. Primary part of the calibration observations presented in this paper were carried out
 during the early science run of the DOT (i.e., 2017A) from 2017 May 11--22. During this period the sky was clear with
 excellent sub-arcsec {\it seeing} conditions ($\sim$0\farcs6--0\farcs9) in the $K$-band, and that is why we could obtain fruitful
 observations in spite of highly humid pre-monsoon weather condition (relative humidity typically $>$70\%) at the
 Devasthal site. Few more additional calibration observations were also performed during 2017 October 7--31, the
 second phase of the early science run (cycle 2017B) after the monsoon. The {\it seeing} during this run was generally
 $\lesssim1''$. However, the best {\it seeing} of 0\farcs45 was noted on 2017 October 16.

The standard NIR observation strategy was followed while observing with the TIRCAM2 which involves acquisition of dark
 frames, along with the observations of flat frames during the morning and evening twilights. In general, short exposure
 observations of the target object in at least 3 dithered positions were deemed sufficient. Typically, in the $JH$
 bands, multiple frames of 50\,s exposure were acquired, while in the $K$-band several 10\,s frames were acquired to avoid
 saturation. Due to large background in the $nbL$-band, several hundreds to thousands of frames were acquired with a short
 exposure of 50\,ms for each frame. Observations of nearby standard stars, immediately before or after the observations
 of science targets, were also done for flux calibration. Additionally, we observed blank sky frames of identical
 exposures as of the target star. This is to remove a non-uniform
 additive illumination in TIRCAM2 images. Master sky frame was generated by median combining all the observed sky frames.
 It is however possible to construct sky frames by median combining the science target frames if the observed field is
 not too crowded. During data reduction, sky background was removed from the science frame by subtracting the master-sky
 images after flat-fielding. We have also constructed a TIRCAM2 bad pixel mask (discussed in Section~\ref{BPmask})
 which was applied to all the science images before performing the photometric reduction.
 
Photometric reduction was carried out using {\sc iraf}\footnote{Image Reduction and Analysis Facility
 (Website: http://iraf.net/)} software, following the standard procedures. All the raw scientific frames were first
 dark subtracted, and then corrected by a master flat. The point sources in those frames were identified using the
 {\sc daofind} package of {\sc iraf} \citep{stetson92}. Aperture photometry was carried out on the frames with isolated
 sources. For the crowded frames, point-spread-function (PSF) photometry was performed. The average PSF was
 defined by choosing at least 9 isolated stars in the frame. Subsequently, the PSF photometry was performed on the
 identified point sources in that frame using the average PSF. The instrumental magnitudes were finally calibrated
 to the standard system using the Two Micron All Sky Survey (2MASS) magnitudes \citep{cutri03} of a few non-variable
 stars in the frame.

\section{Performance of the TIRCAM2 on the DOT}\label{char}
In this section, we present the performance of the TIRCAM2 on the DOT and the estimation of several parameters based
 on the calibration observations.

\subsection{Seeing at Devasthal}
\begin{figure*}
\centering
\includegraphics[width=0.9\textwidth]{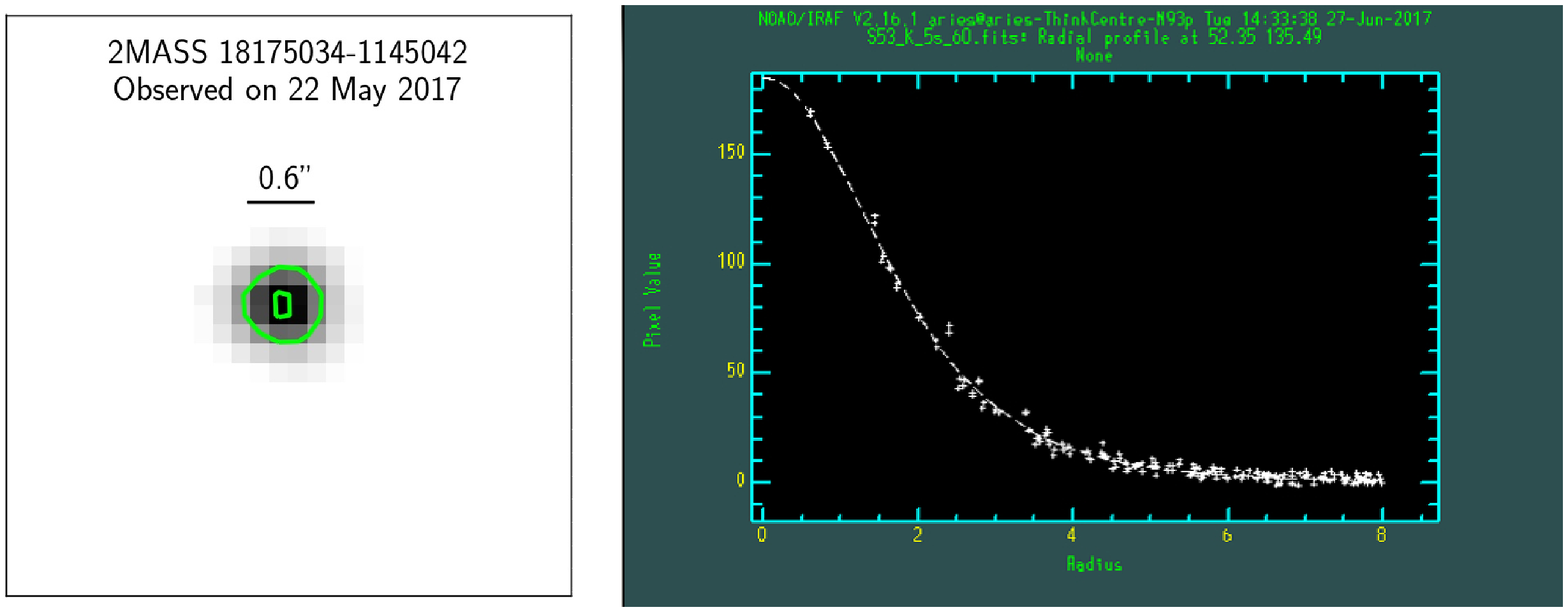}
\includegraphics[width=0.9\textwidth]{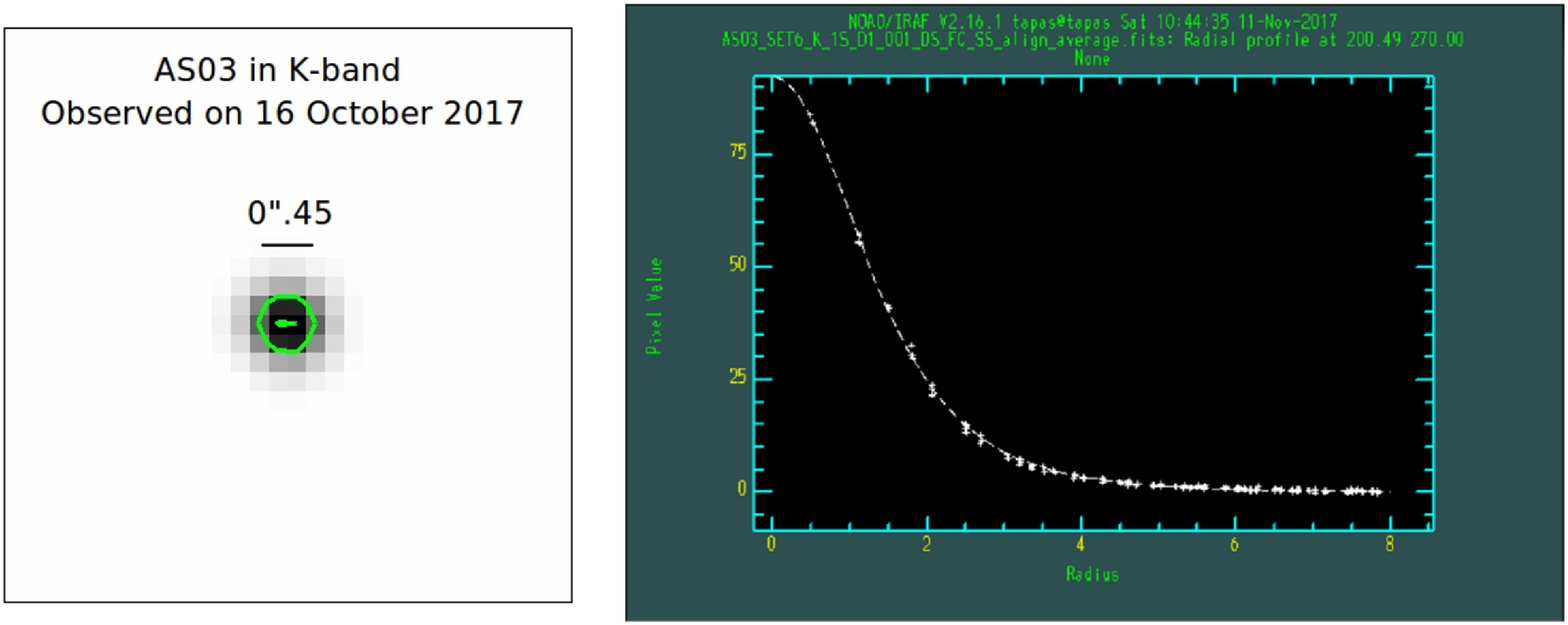}
\caption{{\it Top panels:} Cut-out of a stellar image observed with the TIRCAM2 in the $K$-band on 2017 May 22 towards
 the Serpens OB2 association. Typical {\it seeing} on this night was $\sim$0\farcs6. The outer green contour shows the
 FWHM of the stellar image. The radial profile of the image (in pixel) is also shown in the top-right panel.
 {\it Bottom panels:} A {\it seeing} of $\sim$0\farcs45 was observed on 2017 October 16. The stellar profile of the
 source AS 03 and the corresponding radial profile are shown in the left and right panels, respectively.}
\label{fig2}
\end{figure*}

The FWHM {\it seeing} during the calibration nights was typically sub-arcsecond in all JHK bands. The best {\it seeing}
 obtained during the cycle 2017A was about 0\farcs6 on 2017 May 22 (see a $K$-band stellar image in the top-left panel of
 Figure~\ref{fig2}). We have also shown the radial profile of the stellar image in the top-right panel of Figure~\ref{fig2}.
 The full width at half maxima (FWHM) of the stellar profile is 3.5 pixels which converts to $\sim$0\farcs6 on the sky.
 However, during cycle 2017B, a {\it seeing} of 0\farcs45 ($\sim$ 2.65 pixels) was obtained in the $K$-band on 2017 October
 16, and the corresponding stellar image and the radial profile are also shown in the lower panels of Figure~\ref{fig2},
 respectively. Hence, it seems that under normal sky conditions, the typical {\it seeing} at the DOT site should be
 sub-arcsecond in the NIR bands. Such a sub-arcsecond {\it seeing} is extremely useful not only to resolve sources in
 crowded regions but also to perform accurate photometry. It is also worth mentioning here,
 that the pixel scale of the TIRCAM2 ($\sim$ 0\farcs169) is optimal for sub-arcsecond {\it seeing} conditions at
 Devasthal. It must be mentioned here that the variation of the PSF over the FoV is negligible compared to the
 {\it seeing} limitation.

\subsection{Bad pixel mask}\label{BPmask}
Every detector array has a certain percentage of bad, hot and cold pixels. Bad pixels are considered to be those
 which show substantially high variations with respect to the median counts in the resultant frame obtained by
 dividing two flat frames observed in the high and the low incident flux. For the TIRCAM2, we found that pixels having
 variations more than 8$\sigma$ are generally bad. Similarly, the hot and cold pixels are those pixels that have
 counts more than 8$\sigma$ above and below the median value, respectively, in a dark readout frame. It
 is noted that not all hot and cold pixels are really bad pixels that need to be masked. Efficiency of many of
 these hot and cold pixels can be corrected by dark subtraction and flat field correction.
 
\begin{figure*}
\centering
\includegraphics[width=0.5\textwidth]{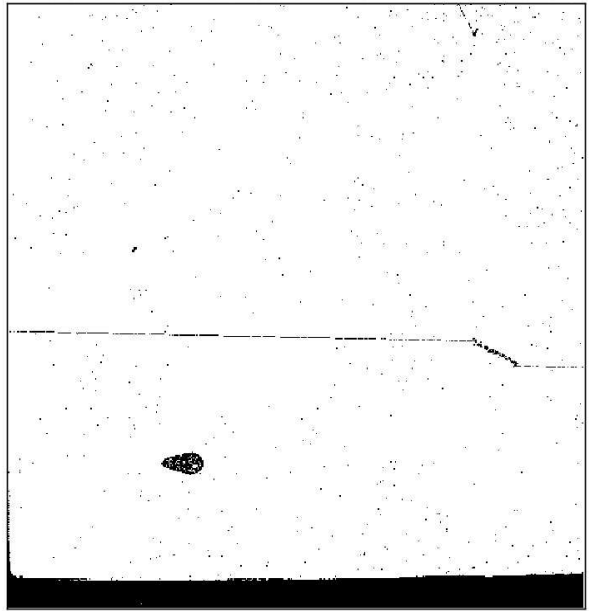}
\caption{Bad-pixel mask of the TIRCAM2 array. Pixels with black dots/patches represent the imperfect pixels in the TIRCAM2
 array that need to be masked.}
\label{fig3}
\end{figure*}

A bad pixel mask has been constructed for the TIRCAM2 array in order to replace the counts in bad pixels by the
 average counts of their neighboring pixels. The bad pixel mask of the array is shown in Figure~\ref{fig3}.
   
\subsection{Sky brightness}
The night sky in the NIR bands is generally a few hundred times brighter than it is in the optical bands. Sky brightness
 in NIR bands depends on various parameters like humidity, presence of the moon and its phase, and ambient temperature.
 The sky brightness values were estimated from the observed science frames after subtraction of dark frames and performing
 flat-field correction\footnote{For detail formula: http://www.inquinamentoluminoso.it/cinzano/en/sbeam2.html}.
 Sky brightness in the $H$, $K$ and narrow $L$ bands ($nbL$) was found to vary night to night. Such
 a variation is also expected since the water vapor contents in the atmosphere were sufficiently high and variable during
 the calibration nights at the DOT. In good sky conditions, we obtained sky surface brightness values (in mag arcsec$^{-2}$)
 of 16.4, 14.0, 12.2 and 3.0 in $J$, $H$, $K$ and $nbL$ bands, respectively. The estimated sky brightness values in both
 the DOT cycles are consistent. Although the sky in the $K$-band at Devasthal seems to be little brighter, but overall
 these values are comparable with other good sites like Hanle \citep{ninan14}, Calar Alto and several other observatories
 \citep[see][]{sanchez08}. The sky brightness in the $nbL$-band also seems to be comparable to the sites like
 Paranal\footnote{https://www.eso.org/gen-fac/pubs/astclim/paranal/skybackground/}.
 
\subsection{Photometric sensitivity and capability}
\begin{figure*}
\centering
\subfigure{\includegraphics[width=0.45\textwidth]{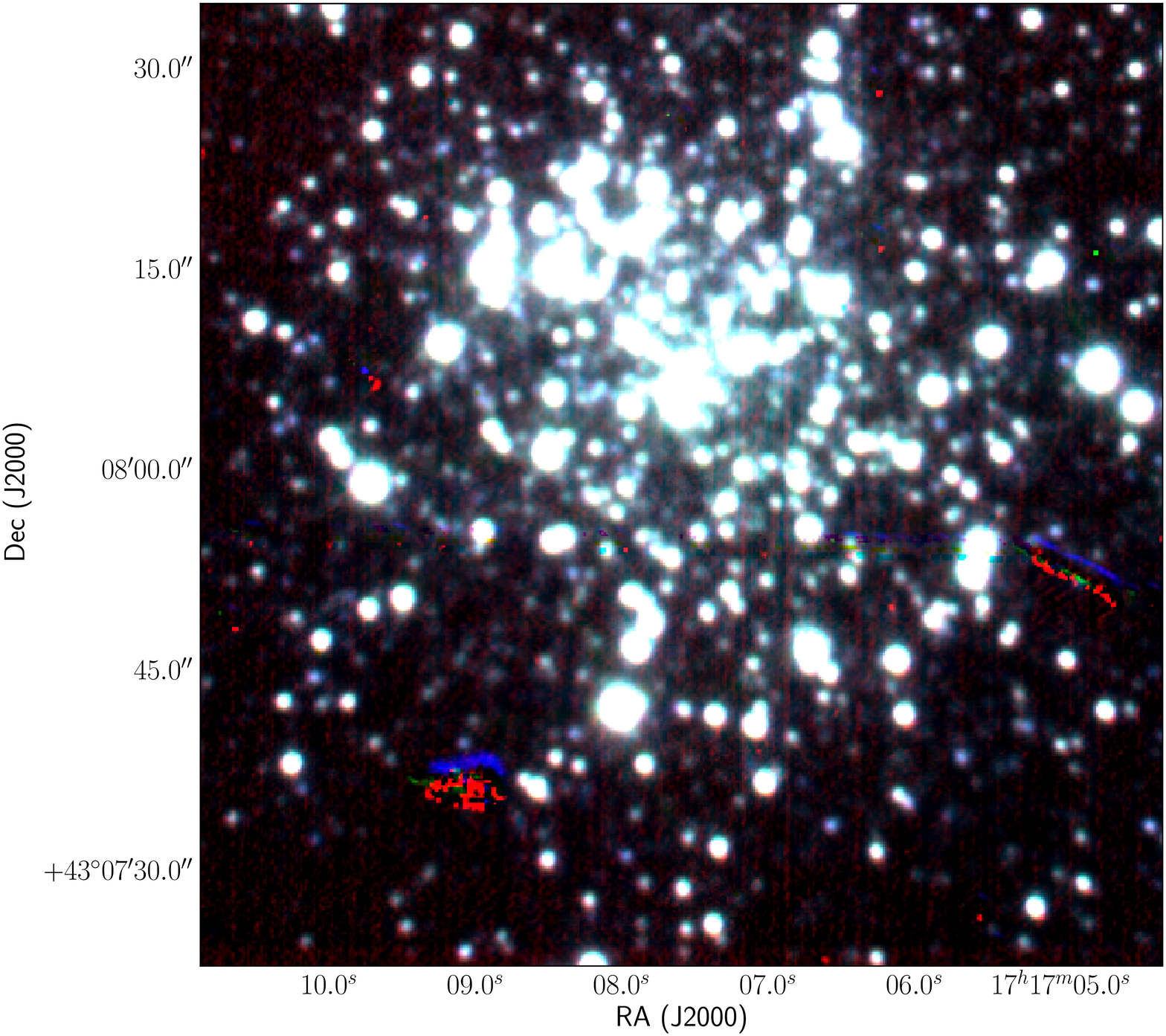}}
\quad
\subfigure{\includegraphics[width=0.45\textwidth]{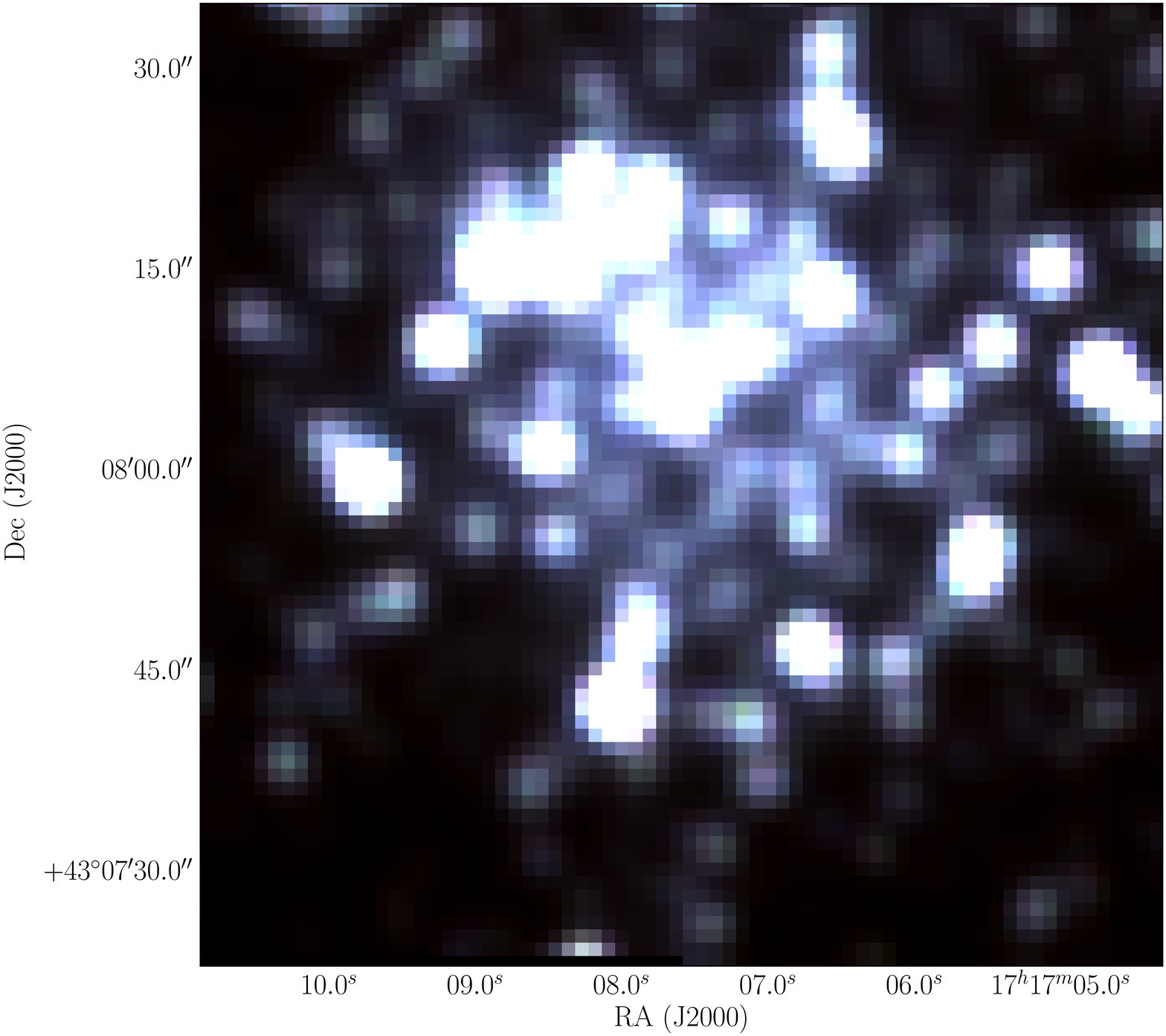}}
\caption{The color-composite image (red: $K$-band, green: $H$-band, blue: $J$-band) of M92, a Galactic globular cluster, constructed
 using the frames observed with the TIRCAM2 attached to the 3.6-m DOT (left). A color-composite image for the same area generated
 using the 2MASS $J$, $H$ and $K$ band images is also shown (right) for comparison.}
\label{fig4}
\end{figure*}

The estimation of the limiting magnitudes in the NIR bands depends on several factors like sky brightness, and reflectivity
 of the primary mirror (i.e., M1). We have observed several clusters to estimate the limiting magnitudes in the NIR $JHK$
 bands. All these clusters were observed in one dithered position in $JHK$ bands. In addition, nearby sky frames were
 also observed with same exposure times as the target frames. Here, we report the limiting magnitudes estimated from the
 globular cluster M92. In both $J$ and $H$ bands, a total of eleven frames were acquired with each frame having an exposure
 time of 50\,s. However, due to large background in the $K$-band, a total of hundred frames were observed with each of them
 having a shorter exposure time of 10\,s. Accordingly, the effective exposure times in $J$, $H$ and $K$ bands are 550\,s,
 550\,s and 1000\,s, respectively. A three-color composite image (red: $K$-band, green: $H$-band, blue: $J$-band) of the
 globular cluster M92 is presented in Figure~\ref{fig4}. In addition, a similar three-color composite RGB image of the
 M92, constructed using the 2MASS $JHK$ bands, is also presented in Figure~\ref{fig4} for comparison. It can be clearly
 seen in the figure that the TIRCAM2 image is much deeper and has a better spatial resolution (FWHM$\sim$0\farcs8) compared
 to the 2MASS images \citep[spatial resolution $\sim$2$''$; see][]{skrutskie06}.

\begin{figure*}
\centering
\includegraphics[width=0.9\textwidth]{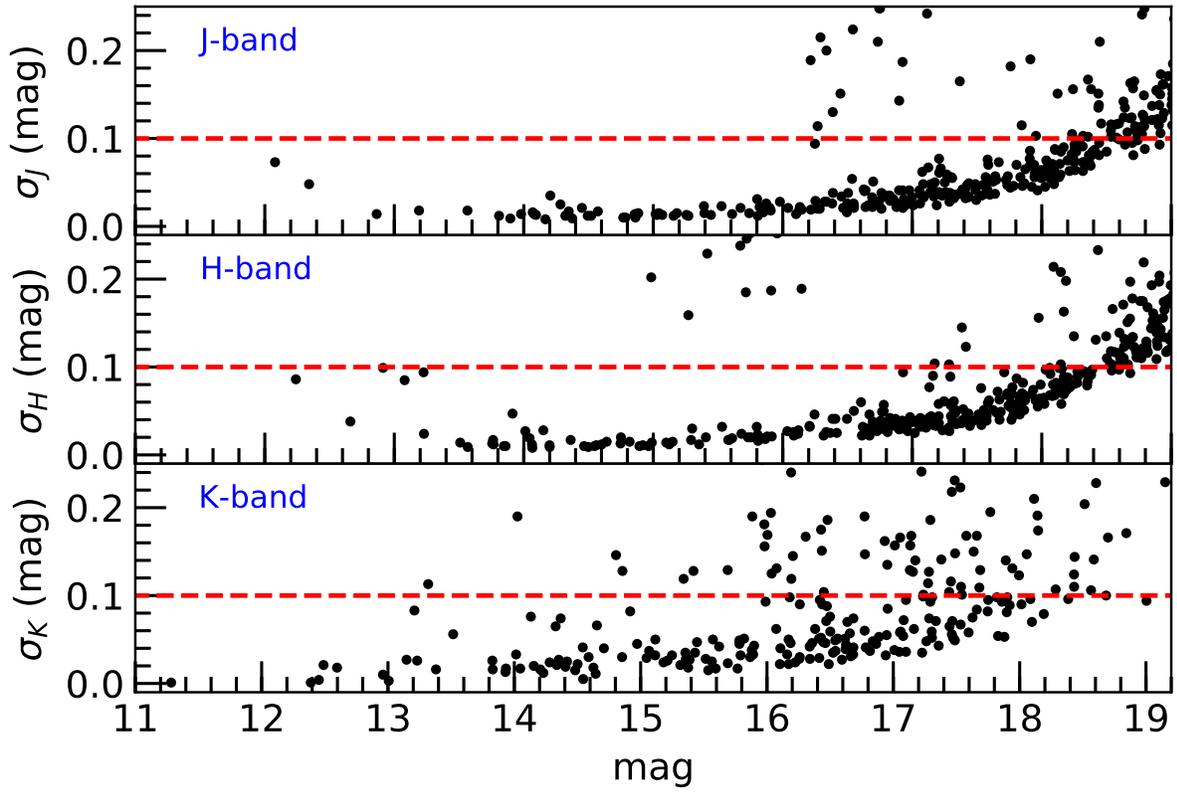}
\caption{$J$, $H$ and $K$ band magnitudes versus magnitude errors of the globular cluster M92 observed with effective exposure times
 of 550\,s, 550\,s and 1000s, respectively. PSF photometry was carried out for frames in all three bands with an aperture
 radius of one FWHM.}\label{fig5}
\end{figure*}

The $JHK$ band magnitudes of the globular cluster M92 are used to estimate sensitivity limits. As can be seen in
 Figure~\ref{fig5}, the 10$\sigma$ limiting magnitudes in $JHK$ bands are of 19.0, 18.8 and 18.0 mag, respectively.
 These numbers are comparable with the limiting magnitudes of other NIR cameras attached to several 4-m class telescopes
 available world-wide, e.g., Wide Field Camera attached to the 3.8-m United Kingdom Infrared Telescope \citep{lawrence07}
 and Wide-field Infrared Camera with the 3.6-m Canada-France-Hawaii Telescope \citep{delorme10}.

\begin{figure}
\centering
\includegraphics[width=0.8\textwidth]{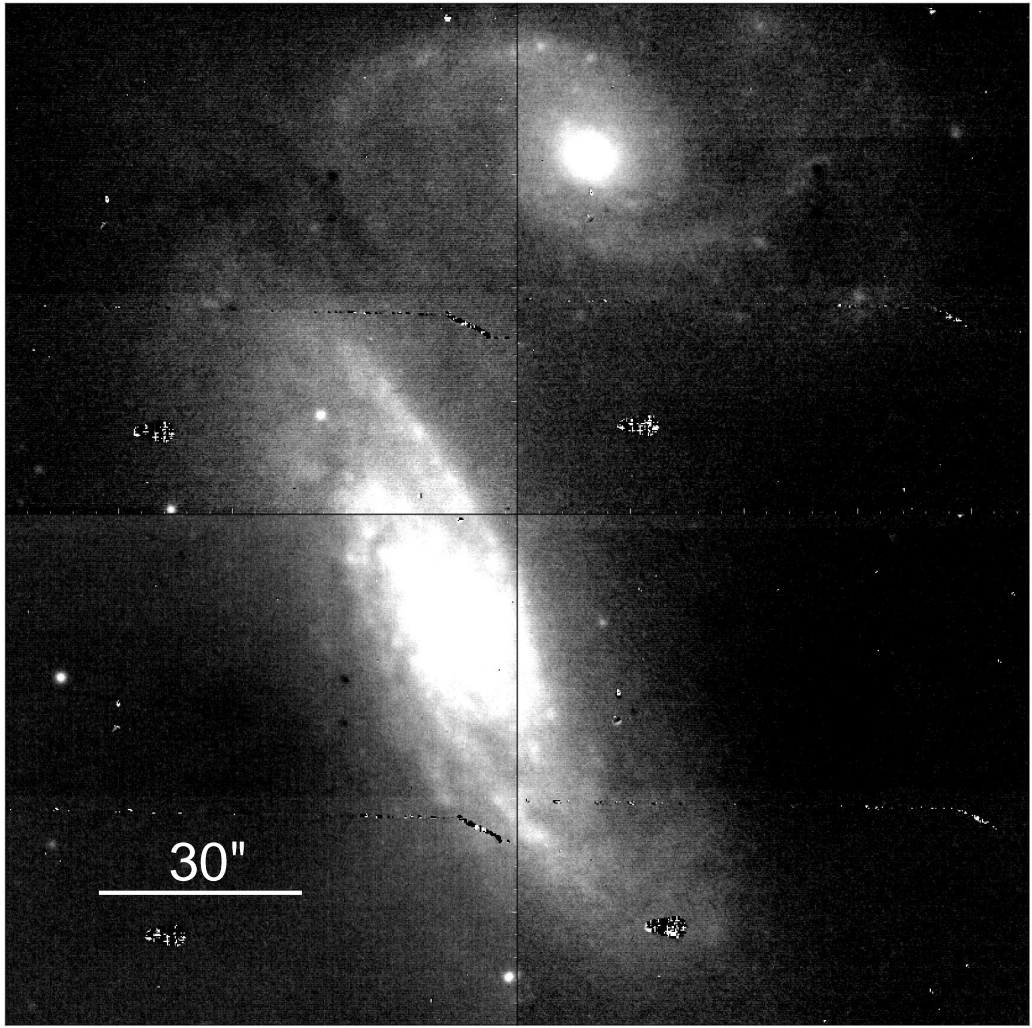}
\caption{Mosaic of four TIRCAM2 $J$-band images (FoV $\sim$ 2\farcm5\,$\times$\,2\farcm5) of twin galaxies, NGC 4567 and NGC 4568,
 observed on 2017 May 15. North is up and East is to left.}\label{fig6}
\end{figure}

The FoV of TIRCAM2 is comparatively small (86\farcs5\,$\times$\,86\farcs5). Thus, in order to examine the dithering capability
 and the linearity of the array, we observed two interacting galaxies, NGC 4567 and NGC 4568, in four dithered positions
 on 2017 May 15. The effective exposure time in each dithered position is 550\,s. A mosaic of four dithered TIRCAM2 $J$-band
 images for a FoV ($\sim$ 2\farcm5\,$\times$\,2\farcm5) covering both the interacting galaxies together is shown
 in Figure~\ref{fig6}.

\subsection{Exposure time calculator}
\begin{figure}
\centering
\includegraphics[width=0.9\textwidth]{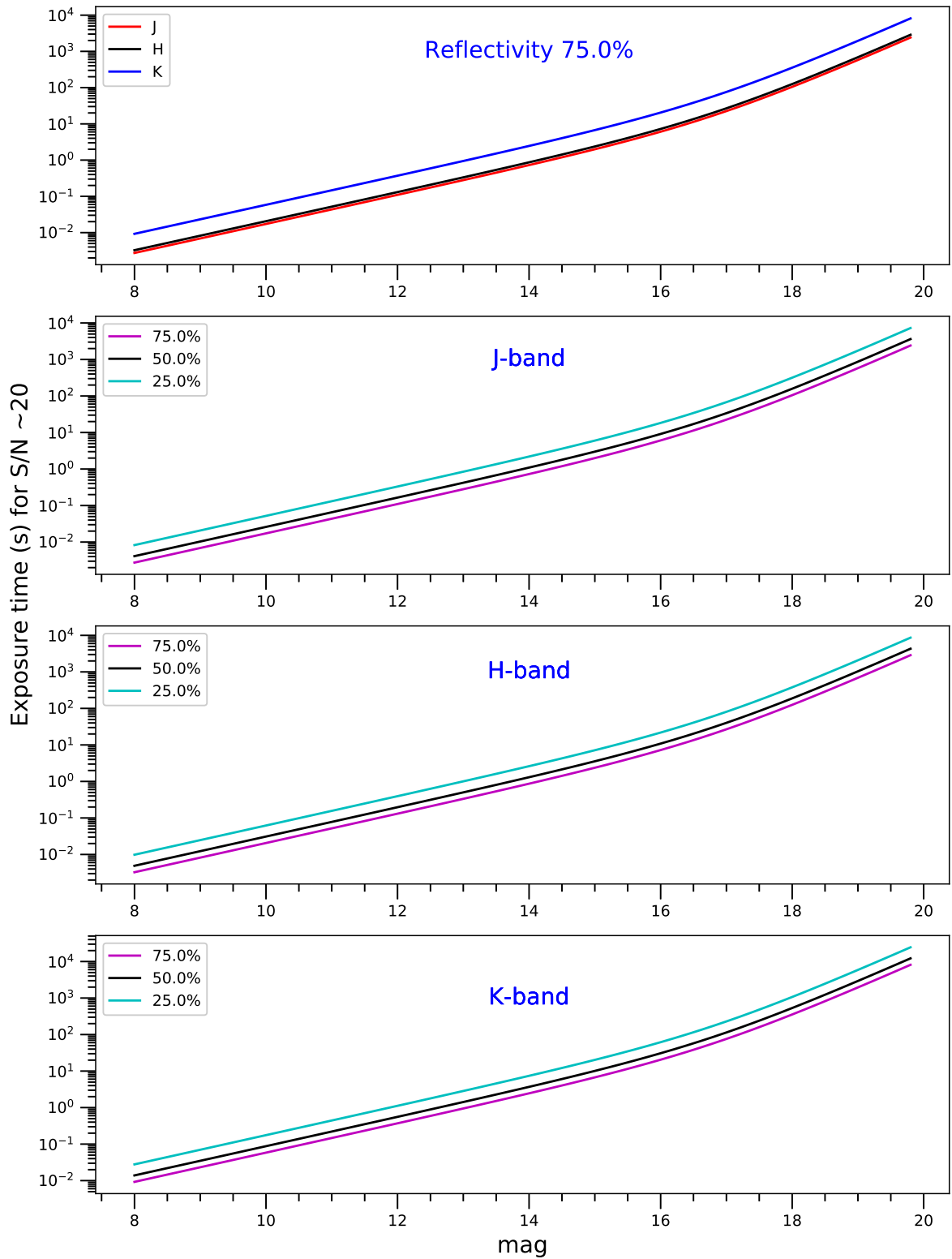}
\caption{Estimated exposure times required for photometry in TIRCAM2 $J$, $H$ and $K$ bands to achieve a S/N$\sim$20 on a typical
 DOT night. The top panel shows sensitivity in all bands with 75\% M1 mirror reflectivity, while other three panels
 show sensitivity for 75\%, 50\% and 25\% M1 mirror reflectivities, separately for each band.}
\label{fig7}
\end{figure}

It is important to estimate the required exposure time for any object before starting observations, as it eventually helps
 to plan and optimize the observations to be made. Hence, we have developed an {\it Exposure time calculator} for TIRCAM2,
 and also made it available online\footnote{http://tirspec.pythonanywhere.com/tircam2/tircam2/}. It requires the name of the
 filter in which observations will be made, magnitude of the target source, expected signal-to-noise ratio (S/N) of the observed
 frame, and the desired number of frames as inputs, in order to obtain an estimate of the required exposure time for
 each frame. It must be noted that per frame exposure times larger than 50\,s, 50\,s and 10\,s in $J$, $H$ and $K$ bands,
 respectively, may saturate the observed frames. In such a situation, the required exposure time has to be re-estimated
 by increasing the number of frames. Figure~\ref{fig7} shows graphs of required exposure times in the $J$, $H$ and $K$ bands
 for a magnitude range from 8--20 mag for a typical 75\% reflectivity of the primary mirror, M1. Estimates of
 required exposure times for three different reflectivities of the M1 (i.e., 75\%, 50\% and 25\%) are also shown in
 Figure~\ref{fig7} separately for all the three bands.

\subsection{Sub-array acquisition mode of TIRCAM2}
\begin{figure*}
\centering
\includegraphics[width=0.9\textwidth]{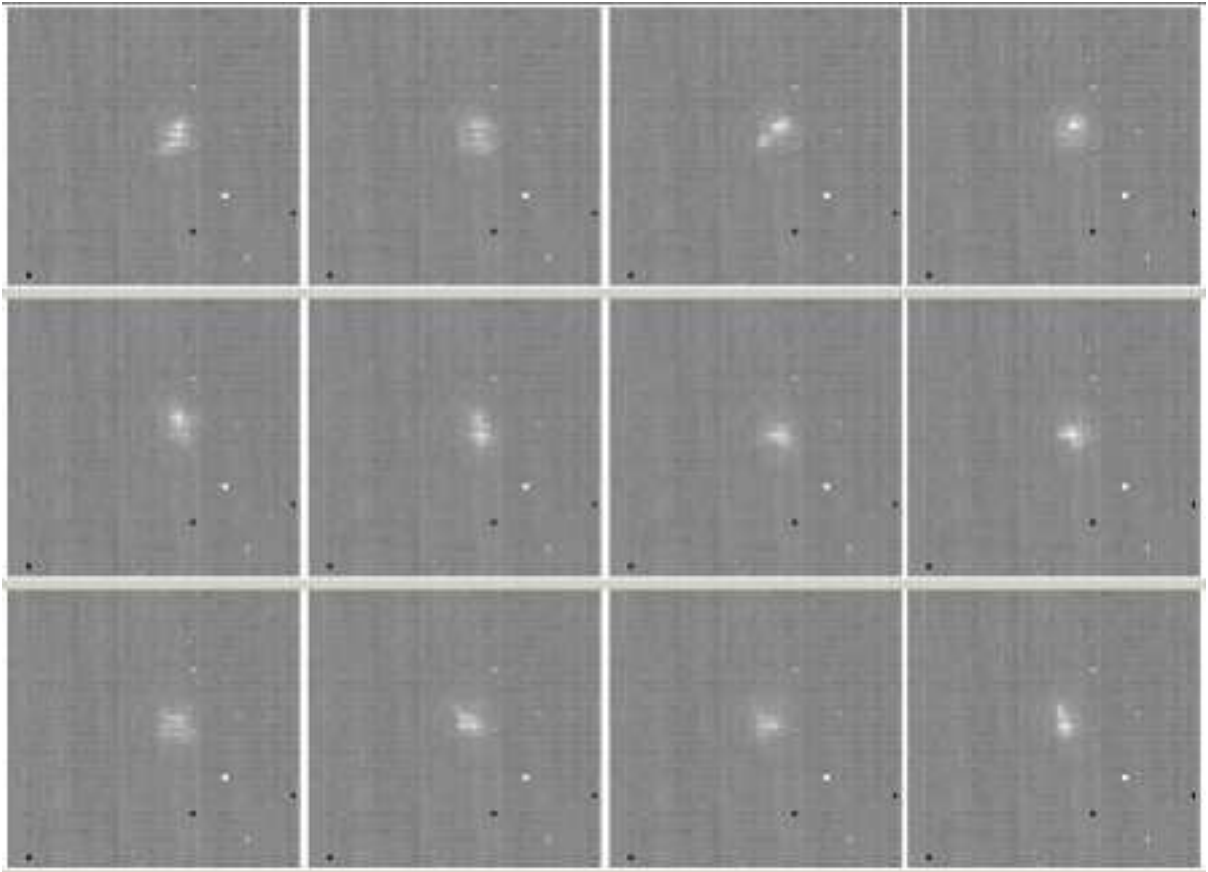}
\caption{Twelve consecutive sky-subtracted frames out of 1000 frames (ascending order row-wise), acquired with
 a window size of 64$\times$64 pixels.}
\label{fig8}
\end{figure*}

In the full frame mode, the TIRCAM2 captures 512\,$\times$\,512 pixels image, and the shortest possible sampling time in
 the full-frame mode is 256 ms. However, in the window or sub-array mode, the TIRCAM2 is able to acquire images for small
 box sizes ranging from 16\,$\times$\,16 to 256\,$\times$\,256 pixels with finer sampling times. This sub-array capability
 of the TIRCAM2 could therefore be useful to observe bright sources with a small exposure time in order to avoid saturation.
 This capability can also be employed for speckle observations generally performed to identify binary stars. Such rapid
 acquisition can also be useful to record light curves of fast events that occur within few hundreds of milli-seconds (e.g.,
 lunar occultations). Twelve consecutive sky-subtracted frames of 64\,$\times$\,64 pixels window (out of 1000 observed frames)
 are shown in Figure~\ref{fig8}. The frame acquisition time for 64\,$\times$\,64 pixels window is $\sim$11 ms and the window
 sampling time is $\sim$8.2 ms. The frame acquisition time is generally more due to the additional time needed to skip rows
 that fall outside the window.

\subsection{Detection of sources in the $nbL$-band}
\begin{figure*}
\centering
\includegraphics[width=0.9\textwidth]{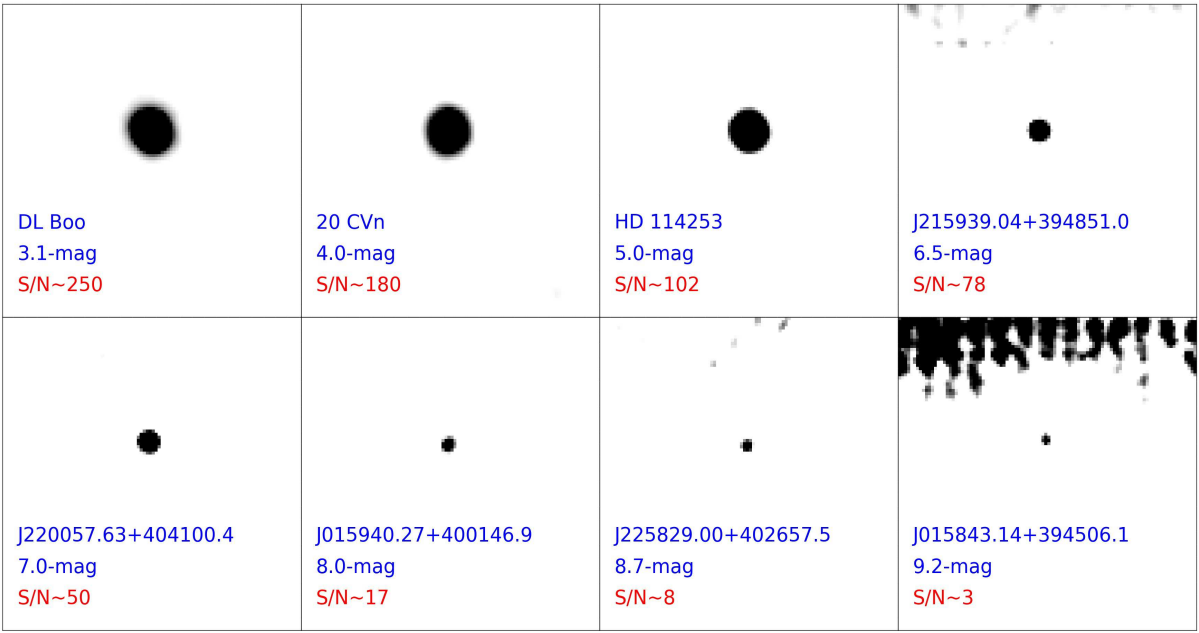}
\caption{Mosaic of the $nbL$-band images (100\,$\times$\,100 pixel cut-outs). Brighter sources (W1~$\lesssim$~5.0 mag) were observed
 on 2017 May 13 with effective exposure time of 15\,s, and the remaining sources were observed on 2017 October 24 with
 effective exposure time of 125\,s. Sources brighter than W1-band magnitude of 6.0 were first aligned and then combined. The
 remaining sources were co-added blindly as these sources are not seen in the frames with exposure time of 50\,ms.}
\label{fig9}
\end{figure*}

Ground-based observations in the NIR $nbL$-band ($\lambda_{cen}\sim$~3.59~$\mu m$, $\Delta\lambda$~0.07~$\mu m$) is extremely
 difficult because of poor atmospheric transmission in this wavelength band. Also, this band is highly affected by the water
 vapor content in the Earth's atmosphere. Observations in this band have primarily been possible with satellite-based cameras,
 such as the {\it Spitzer}-Infrared Array Camera \citep[IRAC;][]{reach05}, and the Wide-field Infrared Survey Explorer
 \citep[WISE;][]{wright10}. We have performed $nbL$-band observations of several sources using the TIRCAM2 attached to the DOT.
 The observed $nbL$-band frames (100\,$\times$\,100 pixels cut-outs) of sources having WISE W1-band (3.4 $\mu m$) magnitudes
 ranging from 3.1 to 9.2 is shown in Figure~\ref{fig9}. The brighter sources (W1~$\leq$~5.0 mag) were observed on the calibration
 nights during the DOT Cycle 2017A with an effective integration time of 15\,s. However, the remaining sources were observed
 during the DOT cycle 2017B with an effective exposure time of 125\,s. It is found that sources up to 9.2 mag can be detected
 in the $nbL$-band of the TIRCAM2 even with an effective exposure time of about 25\,s.

\begin{figure*}
\centering    
\includegraphics[width=0.6\textwidth]{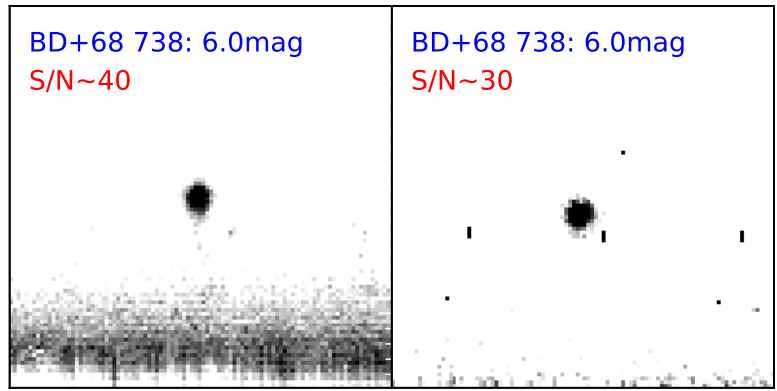}
\caption{The $nbL$-band frames (100\,$\times$\,100 pixel cut-outs) of the source BD+68 738 (W1~$\sim$~5.0 mag) combined after
 alignment (left) and without alignment (right). Both the combined frames have an effective exposure time of 15s. The S/N
 improves if they are aligned and then combined.}
\label{fig10}
\end{figure*}

The sources up to W1-band magnitude of $\sim$6.0 are sparsely visible even in short exposure frames of 50\,ms, and thus,
 it allows us to align and combine those dithered frames to construct the final frame. This eventually helps us to
 achieve a better S/N in the combined frame. However, frames with sources fainter than W1-band magnitude of 6.0 are
 combined without alignment since they were not visible in the individual frames. Note that the S/N of the observed point
 sources can be improved substantially by aligning the frames before combining as compared to combining blindly. For example,
 in Figure~\ref{fig10}, the S/N of the source, BD+68 738 (W1~$\sim$6.0 mag), has improved from 30 to 40 when aligned and
 combined. Hence, it might be possible to observe sources fainter than 9.2 mag with similar exposure in the $nbL$-band using
 the TIRCAM2, if another bright source is present in the frame for alignment. It is worth mentioning here that point sources
 brighter than 7.92 mag and 8.1 mag\footnote{http://wise2.ipac.caltech.edu/docs/release/allsky/expsup/sec6\_3d.html\#satcat}
 are generally saturated in the {\it Spitzer}-IRAC 3.6 $\mu m$ frames \citep{churchwell09}
 and the WISE W1-band images \citep{wright10}, respectively. The TIRCAM2 can therefore be a complementary instrument to observe
 the bright sources that are saturated in the {\it Spitzer}-IRAC and the WISE W1 frames.

\begin{figure*}
\centering    
\includegraphics[width=0.7\textwidth]{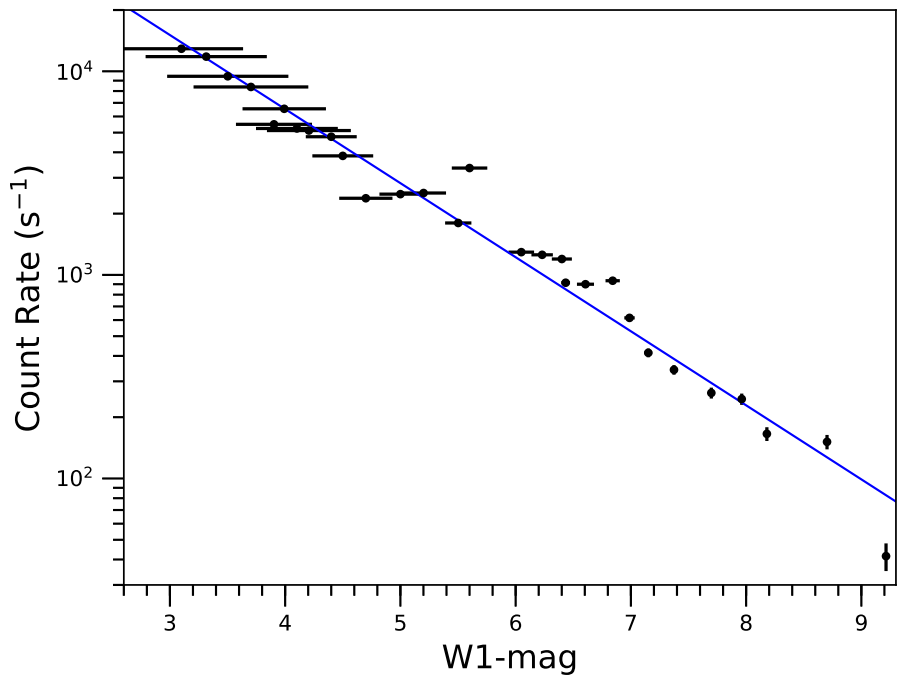}
\caption{The TIRCAM2 count rates $vs$ the WISE W1-band magnitudes ([3.4]), showing the linear behavior of the
 TIRCAM2 in the $nbL$-band in the given magnitude range.}
\label{fig11}
\end{figure*}

We have also examined the linearity of the TIRCAM2 array in the $nbL$-band. A total of 29 sources were observed that have
 W1-band magnitudes ranging from 3.1--9.2. Figure~\ref{fig11} shows the W1-band magnitudes versus corresponding
 TIRCAM2 count rates. As can be seen in the figure, the array behaves linearly in this magnitude range.
 Since the brighter sources in the WISE W1-band ($\lesssim$8.1 mag) are generally saturated, the photometry of these
 sources is determined using an indirect method. Thus, the magnitudes of brighter sources have larger errors.
 However, scatter in the TIRCAM2 count rates in a few measurements are seen possibly because of the variable
 sky background due to high humidity. It must be noted that the source having 9.2 mag is merely a 3$\sigma$ detection,
 and hence, a large scatter (see Figure~\ref{fig11}) in its count is not unusual as the corresponding frame is background dominated.

\subsection{Detection of polycyclic aromatic hydrocarbon emission}
\begin{figure*}
\centering    
\includegraphics[width=0.9\textwidth]{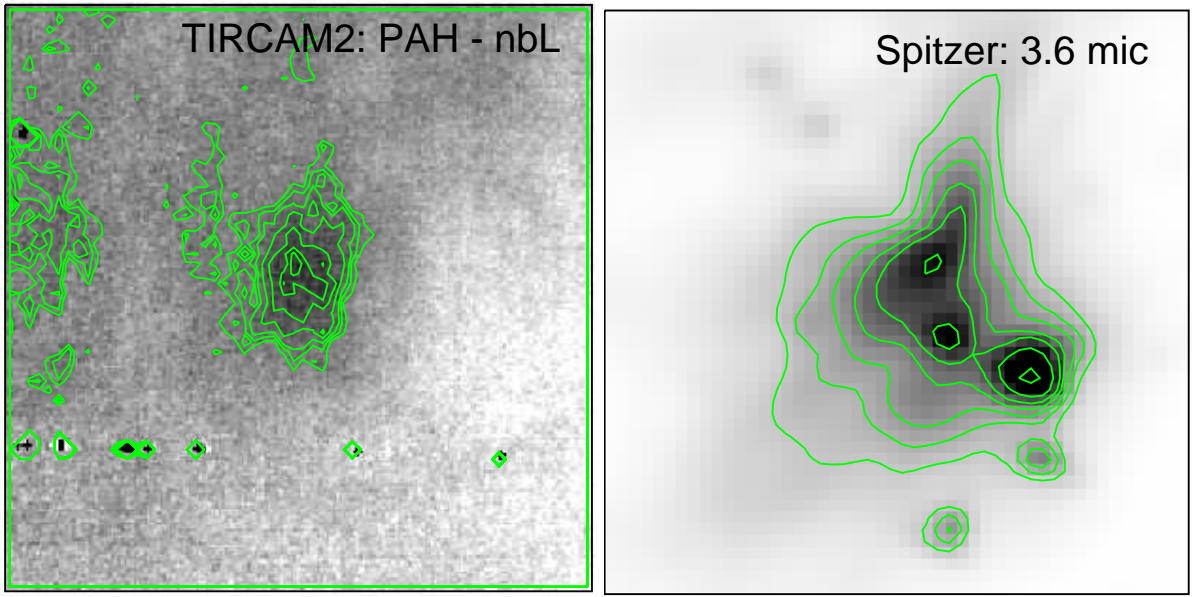}
\caption{The left panel shows the continuum-subtracted $PAH$-band image of 30$^{\prime\prime}$\,$\times$\,30$^{\prime\prime}$
 area around the Sh 2-61 region. Contours are overlaid for clarity. {\it Spitzer} 3.6 $\mu m$ image for the same area is
 also presented for comparison (right panel).}
\label{fig12}
\end{figure*}

The TIRCAM2 is also equipped with a polycyclic aromatic hydrocarbon (PAH) band filter ($\lambda_{cen}\sim$~3.29~$\mu m$,
 $\Delta\lambda$~0.06~$\mu m$). But, in general, it is difficult to observe the diffused PAH emission using ground-based
 telescopes. However, encouraged by the $nbL$-band detection, we observed a few known PAH emitting sources \citep{ghosh02}.
 Figure~\ref{fig12} shows the continuum-subtracted $PAH$-band image of the Sh 2-61 region \citep[F$_{3.3\mu m} \sim$0.4 Jy;
 estimated from][]{verhoeff12} observed using the TIRCAM2 at the 3.6-m DOT with an effective integration time of 13\,s.
 {\it Spitzer}-IRAC 3.6 $\mu m$ image which includes
 PAH emission, is also shown in Figure~\ref{fig12} for comparison. As can be seen in the figure, the TIRCAM2 could detect the
 central (i.e., the strongest) part of the PAH emission of the Sh 2-61 region. Hence, it might be possible to observe strong
 PAH sources (F$_{3.3\mu m} \gtrsim$ 0.4 Jy) using the TIRCAM2.

\section{Summary}\label{summary}
The performance of the NIR imaging camera, TIRCAM2, attached to the 3.6-m DOT is found to be consistent with the expectations.
 At longer wavelengths, specifically at the $nbL$-band ($\lambda_{cen}\sim$~3.59 $\mu m$), the results are highly
 encouraging. In spite of high humidity ($>$70\%) during the calibration observation runs, the {\it seeing} was generally
 sub-arcsecond, and the best {\it seeing} obtained was $\sim$0\farcs45 in the $K$-band on 2017 October 16. Deep imaging
 observations show that the camera has the capability to observe sources up to 19.0 mag, 18.8 mag, and 18.0 mag with 10\% photometric
 accuracy in $J$, $H$ and $K$ bands, respectively, with corresponding effective exposure times of 550\,s, 550\,s and 1000\,s. The
 camera is also capable of detecting the $nbL$-band sources brighter than $\sim$9.2 mag, and hence, can be useful in observing bright
 sources that are saturated in the {\it Spitzer}-IRAC 3.6 $\mu m$ and the WISE W1-band images. Also, the detector response
 is found to be linear in the $nbL$-band up to W1-band magnitude of 9.2. Overall, it is found that the TIRCAM2 with the 3.6-m DOT is
 adequate for deep NIR observations that are comparable to other 4-m class telescopes available world-wide. The camera is also capable of
 detecting strong PAH emitting sources (F$_{3.3\mu m} \gtrsim$ 0.4 Jy), like Sh 2-61. The TIRCAM2 is made available at the
 DOT for scientific observations since 2017 May. It is also proposed to be used on one of the side ports of the DOT as the
 axial port will be occupied by one of the other instruments (e.g., TANSPEC\footnote{TIFR-ARIES Near Infrared Spectrometer},
 ADFOSC\footnote{ARIES-Devasthal Faint Object Spectrograph and Camera}, and CCD imager) in the near future. The corresponding
 mechanical design is being finalized and will be fabricated soon. Overall, it is concluded that the TIRCAM2 can be highly
 useful for deep NIR $JHK$ bands observations and may also be useful for observations in the $nbL$-band, particularly at the
 Devasthal site.

\section*{Acknowledgement}
We thank the anonymous referee for the constructive comments that have improved the presentation of the paper.
 We thank the members of the Infrared Astronomy group of Department of Astronomy \& Astrophysics, TIFR, and specially
 Mr. Rajesh Jadhav and Mr. Shailesh Bhagat for assistance and support during the installation and observations. We also like
 to thank the staff at the DOT, Devasthal and ARIES, for their co-operation during the installation and characterization of
 TIRCAM2. We specially thank Mr. Nandish Nanjappa for his valuable contributions during the installation phase. This publication
 has made use of data products from the Two Micron All Sky Survey (a joint project of the University of Massachusetts
 and the Infrared Processing and Analysis Center/ California Institute of Technology, funded by NASA and NSF).

\begin{wstable}[h]
\caption{TIRCAM2 filter characteristics.}
\begin{tabular}{@{}lcc@{}} \toprule
Filter Name   & $\lambda_{cen}~(\mu m)$ & $\Delta\lambda~(\mu m)$ \\ \colrule
$J$           & 1.20 & 0.36 \\
$H$           & 1.65 & 0.30 \\
$Br$-$\gamma$ & 2.16 & 0.03 \\
$K$-cont      & 2.17 & 0.03 \\
$K$           & 2.19 & 0.40 \\
$PAH$         & 3.28 & 0.06 \\
$nbL$         & 3.59 & 0.07 \\ \botrule
\end{tabular}
\label{table1}
\end{wstable}

\end{document}